\documentstyle[aps,prl,psfig,epsfig]{revtex}
\newcommand{\be}{\begin{equation}}
\newcommand{\ee}{\end{equation}}
\newcommand{\bea}{\begin{eqnarray}}
\newcommand{\eea}{\end{eqnarray}}

\begin{document}
\draft
\title{Orbital Kondo behavior from  dynamical structural defects}
\author{ L. Borda$^{1,2}$,  A. Zawadowski$^{2,3}$, and G. Zar\'and$^{2,4}$,}
\address{
$^1$Ls. f\"ur Theoretische Festk\"orperphysik and Center for Nanoscience, Sektion Physik der LMU M\"unchen, 
Theresienstr. 37, 80333 M\"unchen, Germany,\\
$^2$Department of Physics, and Research Group of the Hungarian Academy of Sciences, \\
Budapest University of Technology and Economics, Budafoki \'ut 8., H-1521 Hungary,\\
$^3$Research Institute for Solid State Physics and Optics, Hungarian Academy of
Sciences, H-1525 Budapest, Hungary \\
$^4$Lyman Physics laboratory, Harvard University, Cambridge MA 02145
}
\twocolumn[\hsize\textwidth\columnwidth\hsize\csname
@twocolumnfalse\endcsname

\date{\today}

\maketitle

\begin{abstract}
{The interaction between an atom moving in a model double-well potential and the
conduction electrons is treated using renormalization group methods in next-to-leading
logarithmic order. A large number of excited states is taken into 
account and the Kondo temperature $T_K$ is computed  
as a function of  barrier parameters.
We find that for special parameters  $T_K$ can be close to  $1 {\rm K}$
and it can be of the same order of 
magnitude  as the renormalized  splitting $\Delta$. However, in the 
perturbative regime we always find that $T_K \alt \Delta $ 
with a $T_K  \alt 1 {\rm K}$ [Aleiner {\em et al.}, Phys. Rev. Lett. {\bf 86}, 2629 (2001)].
We also find that $\Delta$ remains unrenormalized at energies above  the 
Debye frequency, $\omega_{\rm Debye}$.}
\end{abstract}
\pacs{72.10.Fk,72.15.Qm}
]
\narrowtext


\section{Introduction}

There is a great number of experimental anomalies observed by 
point contact\cite{1,2,3},
 dephasing\cite{4,5} and transport\cite{6,7a} measurements in metals at 
low temperature which have
not been theoretically explained in a satisfactory way. 
Sensitivity on annealing\cite{1,5}, 
sample preparation\cite{5}, electro-migration\cite{2} support that   
these anomalies may originate from some 
kind of  dynamical defects.

The  concept of  two-level systems (TLS's), {\em i.e.} atoms or groups of atoms
moving between two positions  resulting in two almost degenerate levels, 
has been introduced to explain the low temperature specific heat anomalies in metallic glasses 
\cite{7}.
In order to keep the level splitting very low, tunneling between the two positions
has been assumed. The TLS model has later been  generalized 
by incorporating  dissipation effects\cite{Leggett} and non-commutative
couplings between the TLS and conduction electrons such as screening
and assisted transition due to the metallic electrons \cite{Kondo,8},  
 and a two-channel Kondo-like behavior (2CK) has been 
conjectured\cite{8}. 
The estimated Kondo temperature $T_K$, however, turned out to be
too small. Taking a few of the higher levels of the  atom into account 
increased $T_K$ considerably\cite{9}, however, this approximation has 
been shown recently to be  incorrect \cite{10,11}: 
Most of the terms cancel out at energies above that of the few lowest levels
if all excited states are included in the computation,
and thus electron 
assisted tunneling results only in a negligible $T_K$ for TLS's \cite{10,11}. 
Similar results were obtained  using the adiabatic approximation for an 
atom moving  in a metallic host\cite{Kagan}.

The physical reason of the above cancellation is that the tunneling 
takes place on a typical time scale $\tau\sim  1/\omega_D$. Therefore 
electrons that are farther from the Fermi surface than $\omega_D$ 
follow the motion of the defect instantaneously, and can be ignored 
when considering complicated correlations between the TLS and the conduction 
electrons.   In other words, the effective bandwidth of the conduction 
electrons is reduced from the Fermi energy $E_F$ down to a value
$\sim \omega_D$.
 
In the present paper we study a similar model but with smaller
or negligible barriers, far from the tunneling regime,
and keeping all excited states through our computations.

We construct the perturbative scaling equations up to next to leading 
logarithmic order and show that 
the cancellation found in Ref.~\onlinecite{10}
also extends to the splitting of the TLS, which remains unrenormalized down to 
the scale  $\omega_{\rm Debye}$ too. This supports again the 
picture of 'adiabatic conduction band', proposed in Ref.~\onlinecite{Kagan}.

In the most interesting cases  the second level is just above or 
around the top of the barrier: In this regime we find  a 
Kondo temperature in the range of $T_K\sim 0.1 -  1{\rm  K}$  using realistic parameters. 
For a special class of parameters the renormalized energy splitting between 
the lowest two levels is around  $T_K$  implying that such dynamical defects 
may give rise to some of the Kondo-like anomalies observed\cite{1,3,6}. 
Our calculations  also indicate that to obtain a measurable 
$T_K$  one needs {\em resonant scattering} on the defect. 

In the present work we only study the perturbative region, and there 
we do not find 
convincing  evidence  of an observable two-channel Kondo behavior.  
However, one can show by performing analytical and numerical 
calculations using a more detailed  model  that there exists a 
regime {\em outside} the range of perturbative calculations 
where the two-channel Kondo behavior appears  \cite{resonant}.

\section{Model}

Our model consists of a particle  (or collective coordinate $z$ of the defect) 
with mass $M$ ($M\sim 50 m_p$; $m_p$ the proton mass) 
moving in some bare potential well, $V_{\rm bare}(z)$.
 The potential $V_{\rm bare}(z)$ can be thought of as the potential resulting
from the interaction with the neighboring  ionic charges. However, 
as we shall see later, $V_{\rm bare}$ is strongly renormalized  due to 
the strong electron-defect  interaction:
In general, both $V_{\rm bare}(z)$ and the 
electron-defect interaction are of the order of the Fermi energy 
($\sim 10 {\rm eV}$),  but their {\em sum}, the effective potential 
is only of the order of the Debye frequency.  Therefore  $V_{\rm bare}(z)$
has very  little physical meaning. 

The non-interacting part of the Hamiltonian is given by:
\be
H_0=\sum\limits_n \varepsilon_nb^{\dagger}_nb_n+
\sum\limits_{{\bf p},\sigma}\varepsilon_{\bf p} a^{\dagger}_{{\bf p},\sigma}a_{{\bf p},\sigma}\;.
\label{H0}
\ee
The first term describes the motion of the  particle in the potential $V_{\rm bare}$:  
$b^\dagger_n$ ($n=1,2,\dots$; $\sum_n b^\dagger_n b_n \equiv1$ ) 
denotes the creation operator corresponding to a state of the particle with 
energy $\varepsilon_n$ and  wave function $\varphi_n(z)$. 
The second term describes the conduction electrons, 
$a^{\dagger}_{{\bf p},\sigma}$ being the creation operator of an electron 
with momentum ${\bf p}$, spin $\sigma$, and  energy $\varepsilon_{\bf p}$.
In the following we shall use a simple free electron approximation 
to describe the conduction band, and assume that the wave function
corresponding to $a^{\dagger}_{{\bf p},\sigma}$ is a simple plane wave
$\sim e^{i {\bf k\; r}}$.
The density of states for the electrons is 
$\varrho(\varepsilon)\approx \varrho_0(1+\alpha\varepsilon/D_0)$, where $\varrho_0$ is
the value at the Fermi energy $\varepsilon=0$, $D_0$ is a symmetric bandwidth
cut-off of the order of the Fermi energy, and $|\alpha|<1$ accounts for possible electron-hole symmetry
breaking due to 
band structure effects.

We use a simple local interaction between the particle and the electrons:
\bea
H_{\rm int}=\sum\limits_\sigma\int dz\;dz_p\Big[\psi^{\dagger}_\sigma(z)
\psi_\sigma(z)\Psi^{\dagger}(z_p)\Psi(z_p)\nonumber\\
\left(U_0\delta(z)+U(\delta(z-z_p)-\delta(z))\right)\Big]\;,
\label{H1}
\eea
where $\psi_\sigma (z)$ is the electron field operator along the axis $z$ 
of the defect motion ($x=y=0$) and $\Psi(z_p)= \sum_n b_n\; \varphi_n(z_p)$ is the particle field 
operator at the position of the heavy particle, $z_p$. $U_0$ describes a static scattering  at the
center
while $U$ describes the change in the scattering potential when the particle
is displaced from the center. 

In the following we shall simply take $U_0=0$. This simplification 
needs some explanation: In principle, the value of $U$ should be precisely 
$U_0$ in Eq.~\ref{H1}. One can, however, argue, that the electronic wave functions 
corresponding to the operators $a^{\dagger}_{\bf p}$ in Eq.~\ref{H0}   
should be determined when the particle is at the origin, and should therefore already 
incorporate the effect of $U_0$, which could therefore 
be ignored.

The previous argument is, however, not quite right. The consequences of the presence of $U_0$ 
would be quite important. The term $\sim U_0$ can be treated exactly for a single impurity
\cite{14}:
Since $U_0$ describes a static potential it can be
incorporated in the electronic wave functions, which cannot be approximated by plane waves 
any more.
Its most important effect is to  {\em renormalize} the local density of states
$\varrho_0$ in the $s$-channel 
as $\varrho_0 \to {\rm cos}^2(\delta_0)\;\varrho_0$,
where $\delta_0$ denotes the scattering phase shift induced by the 
scattering potential $U_0=U$. As we shall see below, experiments indicate that the 
value of $U$ is rather large, and corresponds to almost  resonant scattering. 
In other words, in this simple model with $U= U_0$ at  resonant scattering 
$\delta_0 \approx \pi/2$, and the local density of states is strongly 
suppressed.  As a result, the dimensionless couplings estimated below are 
reduced. This is a serious problem: One can show that for 
a potential scattering model with $U=U_0$ this suppression is so strong
that one is always in the weak coupling regime,  and one can therefore 
never observe the two-channel Kondo behavior, while without the term $U_0$ 
one can go beyond the Emery  - Kivelson line and prove even analytically 
the existence of a regime where a two-channel Kondo behavior appears 
\cite{resonant}.   

To resolve this problem, one has to go beyond approximating the particle 
by a simple potential scatterer and one must take into account
the dynamics of the internal electronic degrees of freedom 
of the particle \cite{resonant}. This analysis turns out to justify
our approximation of setting $U_0=0$ in Eq.~(\ref{H1}).

To handle the interaction part of the Hamiltonian we follow Ref.~\cite{8} 
and  introduce spherical coordinates. Assuming that the motion of the defect 
is restricted in space ($k_Fz_p < 1$, with $k_F$ the Fermi momentum), the dominant 
electron scattering  occurs in the  ($l=0,m=0$) and ($l=1,m=0$) angular 
momentum channels, and we can neglect scattering channels with $l>1$ \cite{8}. 
$H_{\rm int}$ then becomes:
\be
H_{\rm int}={1\over{2\pi}}\sum\limits_{n',n\atop l',l,\sigma}
\int dk\int dk'\; V^{n'n}_{l'\;l}b^{\dagger}_{n'}b_na^{\dagger}_{k'l'\sigma}
a_{kl\sigma}\;,
\label{H1mod}
\ee
where $a^{\dagger}_{kl\sigma}$ creates an electron with angular momentum 
$l$ ($m=0$), radial momentum $k$ and spin $\sigma$, and satisfies the 
anti-commutation relation $\{ a^{\dagger}_{kl\sigma}, a_{k'l'\sigma'}\}
= 2 \pi\; \delta(k-k')\; \delta_{\sigma\sigma'} \;\delta_{l l'}$.
We evaluated the dimensionless interaction matrix elements $V^{n'n}_{l'\;l}$ using
spherical wave functions with momentum $k\approx k_F$ for the electrons 
and the exact wave functions $\varphi_n$ for the defect. 
The dynamics of the heavy particle is described in terms of the 
pseudo-fermions $b^{\dagger}_{n}$ satisfying the constraint 
$\sum_n b^{\dagger}_{n}b_{n} = 1$.

\section{Renormalization Group Analysis}

\begin{figure}
\begin{center}
\epsfxsize7.0cm
\epsfbox{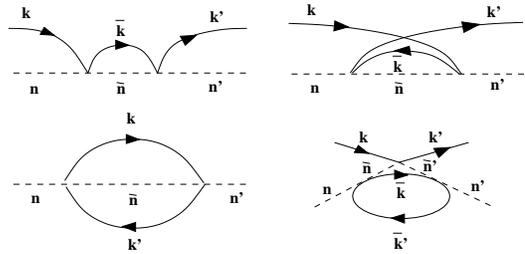}
\end{center}
\caption{\label{diagrams} 
Diagrams used to construct the renormalization group equations:  
Dashed lines correspond to  the defect atom, solid lines to the electrons.
}
\end{figure}

To determine $T_K$ we  constructed the leading and  next-to-leading logarithmic 
renormalization group (RG) equations  \cite{9}.  To this end one has to compute 
vertex and self-energy corrections to the pseudo-fermion propagator 
${\cal G}_{nn'}(\tau) \equiv - \langle T_\tau b_n(\tau) b_{n'} (0)\rangle$
and the impurity-electron vertex function $\Gamma^{nn'}_{ll'}$, and then apply the 
relatively standard machinery of multiplicative RG. 
The corresponding skeleton diagrams are shown in Fig.~\ref{diagrams}. 
[Note that the first order self-energy diagram (not shown in Fig.~\ref{diagrams}) 
only generates a local non-logarithmic and time-independent 
renormalization of the defect potential and can therefore be  entirely ignored.]
In the RG procedure the bandwidth cut-off is reduced $D\rightarrow D+dD$ and 
the couplings are simultaneously changed to keep physical quantities 
invariant.  In the present case a matrix  version of the multiplicative RG 
must be employed \cite{zar96}.  

In this work  we take also into account  the energy dependence of the local density of 
states. This may also depend  on the angular momentum channel $l$,  
$\varrho_l(\varepsilon)/\varrho_l(0)=1+\alpha_l\varepsilon/D_0$, where the slope 
$\alpha_l$ and value $\varrho_l(0)$ of the density of states is usually different 
for different values of $l$. The energy $\epsilon$ is measured from the Fermi energy 
and only the linear term has been kept in $\varrho_l$.

\begin{figure}
\begin{center}
\epsfxsize7.0cm
\epsfbox{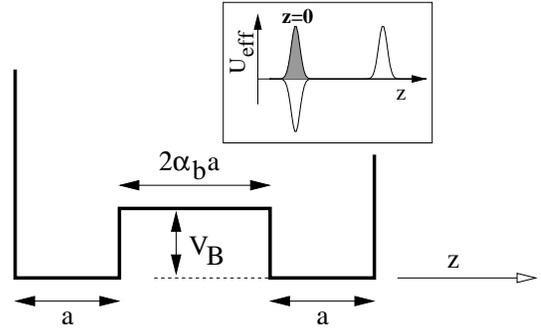}
\end{center}
\caption{\label{tower} 
The simple square-shaped double-well potential used:
$V_B$ is the height of the barrier, $a\sim 0.1{\rm\AA}$ 
is the width of the 
wells and $2\alpha_ba\sim 0.4-0.5{\rm\AA}$ 
is the barrier width. Inset: Effective scattering potential for the electrons. The shaded
potential corresponds to the defect sitting at the center.  The white
potential is due to the displacement of the defect.
}
\end{figure}

Let us first focus on the case of $\varrho_{l=0} = \varrho_{l=1}=\varrho_0$ and $\alpha_0 
= \alpha_1=\alpha$. 
Then the second order self-energy correction of the pseudo-fermion propagator, 
e.g., contains both logarithmic and non-logarithmic corrections, and is given by the following 
expression:
\begin{eqnarray}
\Sigma_{nm}(\omega)  =  
&& - 2 \sum_{\epsilon_{\tilde n} < D_0} {\rm tr} \{{\underline V}^{n \tilde n} 
{\underline V}^{\tilde n m}\}
\label{eq:sigma}
\\
\;
&&\Bigl[ D_0 \; f(\alpha) + 
 (\omega - \epsilon_{\tilde n}) 
\bigl\{\ln {D_0 \over |\omega - \epsilon_{\tilde n}|} + g(\alpha)  \bigr\} \Bigr]\;, 
\nonumber 
\end{eqnarray}
where we introduced a matrix notation, 
$V_{l\; l'}^{n  n'} \to {\underline V}^{n n'} $, 
 ${\rm tr}\{ ...\}$ denotes the trace in the electronic angular momentum indices,
and  the constants $f(\alpha)$ and $g(\alpha)$ are given by 
\begin{eqnarray}
f(\alpha) & = & {2\over 3} \alpha^2 (\ln 2 -1 ) + 2 \ln 2\;, \\
g(\alpha) & = & 1 - \ln 2 + \alpha^2 (\frac12 - \ln 2) \;.
\end{eqnarray}
The factor two in Eq.~(\ref{eq:sigma}) is due to the spin degeneracy of the electrons.
The effect of the non-logarithmic terms proportional to $f$ and $g$ is to strongly  renormalize 
the eigenstates and eigenfunctions of the double well potential by replacing 
the  the heavy particle Hamiltonian by
\begin{equation} 
\sum\limits_n \varepsilon_nb^{\dagger}_nb_n \to 
\sum\limits_{n,m} \Delta_{nm}b^{\dagger}_nb_m
 \end{equation}
 where the matrix $\Delta_{nm}$ is given by 
\begin{equation}
\Delta_{nm} = 
\varepsilon_n \delta_{nm} 
- 2 \sum_{\tilde n} {\rm tr} \{{\underline V}^{n \tilde n} 
{\underline V}^{\tilde n m}\}
\bigl[ D_0 \; f(\alpha) - \epsilon_{\tilde n}  g(\alpha)  \bigr]\;.
\end{equation}
The effect of the static ($\omega$-independent) non-logarithmic terms in Eqs.~(\ref{eq:sigma}) 
can be taken into account by  diagonalizing $\Delta_{nm}$   through a {\em unitary 
transformation} 
\begin{eqnarray}
(U\Delta U^\dagger)^{nm}  = {\tilde \epsilon}_m \delta_{nm}\;,\\
{\underline {V}}^{nm} \to  (U {\underline {V}}
U^\dagger)^{nm}\;,
\end{eqnarray}
where the ${\tilde \epsilon}_n$'s denote the renormalized 
values of the heavy particle  energies. The effective Hamiltonian corresponding 
to $\Delta_{nm}$ generally contains non-local terms too, but the largest terms actually 
turn out to be simple local corrections to $V_{\rm bare}$. 
Therefore, in the rest of the paper we shall drop these non-logarithmic 
self-energy corrections, and assume instead that we can model the entire effective Hamiltonian 
by a simple local  quasi one-dimensional symmetrical square potential with a barrier in the
middle and  infinite walls (see Fig.~\ref{tower}). We determine the
corresponding wave functions by solving simple transcendental equations
and then use these wave functions to compute the interaction matrix elements
${\underline {V}}^{nm}$.

Note that it is only the renormalized Hamiltonian that can be measured and has therefore 
physical meaning. Note also, that both $V_{\rm bare}$ and the the corrections 
are usually of the order of the Fermi energy, however, their {\em sum} is usually 
of the order of the Debye energy only.

The interaction vertex also contains a non-logarithmic 
part. Up to second order in ${\underline V}$ the dimensionless 
vertex function can be calculated as
\begin{eqnarray} 
{\underline  \gamma}^{nm}(\omega) &=& 
{\underline  V}^{nm}
- \alpha \sum_{\epsilon_{k} < D_0}
 \bigl\{
{\underline V}^{nk} {\underline V}^{k m} +
{\underline V}^{k m} {\underline V}^{nk} 
\bigr \} \nonumber \\
&- & 
\sum_{\epsilon_{k} < D_0} 
\ln{D_0\over|\omega-\epsilon_k|} 
\big [
{\underline V}^{nk}, {\underline V}^{k m}]\;.
\label{eq:vertex}
\end{eqnarray}

The effect of the non-logarithmic terms in   (\ref{eq:vertex}) can be taken into account 
by renormalizing the bare vertex as 
\begin{equation}
{\underline  V}^{nm} \to {\underline {\tilde  V}}^{nm}  \equiv {\underline  V}^{nm} 
- \alpha \sum_{k} \bigl\{
{\underline V}^{nk } {\underline V}^{k m} +
{\underline V}^{k m} {\underline V}^{n k}
\bigr \}\;.
\end{equation}
This transformation must be constructed order by order,  and it sums up 
systematically all higher order non-logarithmic vertex contributions.  

After performing the transformations above the expressions of ${\cal G}^{-1}$ and $\gamma$ simplify 
considerably, and  contain only logarithmic terms in $D_0$:
\small
\begin{eqnarray}
[{\cal G}^{-1}]_{nm}(\omega) &=&  
\omega -\delta_{nm} {\tilde \epsilon}_n 
\label{eq:sigma2} \\
&+&  2 \sum_{{\tilde \epsilon}_{k} < D_0} 
{\rm tr} \{{\underline {\tilde V}}^{n k} 
{\underline {\tilde V}}^{k m}\}
 (\omega - {\tilde \epsilon}_{k}) 
\ln {D_0 \over |\omega - {\tilde \epsilon}_{k}|}\;, 
\nonumber 
\\
{\underline  \gamma}^{nm}(\omega) &=& 
{\underline  {\tilde V}}^{nm}
-   \sum_{\epsilon_{\tilde n} < D_0} \ln{D_0\over|\omega-\epsilon_n|} 
\big [
{\underline {\tilde V}}^{n\tilde n}, {\underline {\tilde V}}^{\tilde n m}]\;.
\label{eq:vertex2}
\end{eqnarray}
\normalsize

The remaining logarithmic terms in 
Eqs.~(\ref{eq:sigma2}) and (\ref{eq:vertex2}) 
can be summed up using a generalized version of the 
 multiplicative RG \cite{zar96}, leading to the following RG equations:
\bea
{{\rm d}{\underline {\tilde V}}^{ij}\over{{\rm d}x}} & = &
-\sum\limits_{{\tilde \epsilon}_k< D } \bigl 
[{\underline {\tilde V}}^{ik},{\underline {\tilde V}}^{kj}\bigr]
+
\sum\limits_{{\tilde \epsilon}_k,{\tilde \epsilon}_l< D}
 \; 2\; \Bigl( 
{\underline {\tilde V}}^{kl}{\rm tr}\big[{\underline {\tilde V}}^{ik}{\underline {\tilde V}}^{lj}\big]
\nonumber \\
&
-  & \frac12\; {\underline {\tilde V}}^{ik}{\rm tr}\big[{\underline {\tilde V}}^{kl}{\underline {\tilde V}}^{lj}\big]
- \frac12 \;{\rm tr}\big[{\underline {\tilde V}}^{ik}{\underline {\tilde V}}^{kl}\big]{\underline {\tilde V}}^{lj}\Bigr)\;.
\label{renorm_eq}
\eea
Here $x={\rm ln}(D_0/D)$ is the
scaling variable, and $D_0$ denotes the initial 
value of the cut-off. 
As the Hamiltonian is diagonal in the spin,  each closed electron
loop results in a multiplicative factor of 2 corresponding to 
${\mathcal N}_f=2$ conduction electron channels.

The renormalization group equations for the energies ${\tilde \epsilon}_n$
are somewhat more complicated. In particular, the RG generates off-diagonal 
terms to the heavy particle Hamiltonian:
\begin{eqnarray}
\delta_{nm} {\tilde  \epsilon}_n &\to& 
\delta_{nm} {\tilde  \epsilon}_n 
+ {\rm d}x \sum\limits_{{\tilde \epsilon}_k< D}\Bigl \{   
 2
{\tilde \epsilon}_k{\rm tr}\big[{\underline {\tilde V}}^{nk}{\underline {\tilde V}}^{km}\big]
\nonumber \\
&{ }&\qquad\;{\tilde \epsilon}_n{\rm tr}\big[{\underline {\tilde V}}^{nk}{\underline {\tilde V}}^{km}\big]
- {\rm tr}\big[{\underline {\tilde V}}^{nk}{\underline {\tilde V}}^{km}\big] {\tilde \epsilon}_m
\Bigr \}\;,
\end{eqnarray}
where ${\tilde \epsilon}_n,{\tilde \epsilon}_m < D $, and ${\rm d}x = \ln [D/D']$. 
Therefore,
in each RG step  we diagonalize the self-energy by constructing a new set of 
defect states  with renormalized energy eigenvalues, and express 
all couplings in this new basis: $\tilde\varphi_n = \sum_m U_{mn} \varphi_m$. 
Note that Eqs.~(\ref{renorm_eq}) are invariant under this 
unitary transformation. 

When the  reduced cutoff 
reaches the renormalized energy eigenvalue of some defect level,
the dynamics of that level is frozen 
out, and we therefore drop it in the following RG steps. 
Usually only a few levels remain active in the region where the relatively 
weak initial couplings become comparable to the stronger ones. 
In many cases  only one level remains by this time, meaning that no Kondo 
effect occurs at all.

In the original tunneling model\cite{8} 
only the states $n=1,2$ were kept giving rise to  
two initial dimensionless coupling constants, $v^x$ and $v^z$, corresponding to the
diagonal and off-diagonal terms in indices $n$ and $n'$:
\begin{eqnarray}
v^z={1\over 4} (V^{21}_{01}+V^{21}_{10}+V^{12}_{01}+V^{12}_{10})\;,
\label{eq:vz}
\\
v^x={1\over 4} (V^{11}_{00}-V^{11}_{11}-V^{22}_{00}+V^{22}_{11})\;.
\end{eqnarray}
We identify Kondo temperature $T_K$  with the energy scale
at which the initially
small dimensionless coupling $v^x$
reaches 
 about the half of 
its fixed point value,   $v_x\sim 0.1$ \cite{footnote}, and we keep 
the last two levels active even if their separation is larger than 
the running cut-off $D$.  Below the Kondo
temperature the perturbative RG breaks down and more efficient methods 
are needed. If only two levels are kept then Bethe-ansatz\cite{16}  
results for the  two-channel Kondo problem could be applied. 

The bare  values of  $v^z$ can be estimated  from ultrasound data \cite{8} 
and for alloys with stronger couplings they
correspond to $v^z \sim 0.2$ and thus $U\varrho_0\sim 3$ 
for typical parameters. 
 {This rather large value of 
$U\varrho_0$ implies {\em resonant scattering} on the atom \cite{footnote2}.
Nonetheless, even  $U\varrho_0\sim 3$ turns out to be 
 small enough so that the couplings $V^{nn'}$ corresponding  to it are 
still in the perturbative regime:}
Restricting our discussion to  $n=1,2$, the largest matrix elements are those where
both  the electron and defect parities are changed and  are about $v^z \sim0.2$.  
The reduction from $U\varrho_0\sim 3$ is due to a factor $v^z \sim k_F d\; U\varrho_0$, 
with $d= 2a(1+\alpha_b) \sim 0.4 \; {\rm \AA}$  the width of the potential well, 
and $k_F^{-1}\sim 1{\rm \AA}$ the Fermi momentum. Couplings with conserved 
parities such as $v_x$ are even  smaller, and  are  further reduced by the Gamow factor 
in the case of a large barrier in the tunneling regime \cite{8}.

\section{Results}

In this work we used  the following procedure: We first diagonalized 
the effective heavy particle Hamiltonian to obtain the renormalized 
defect energies ${\tilde \epsilon}_n$ and the renormalized couplings 
${\tilde V}$ numerically.  
Then we summed up logarithmic terms by performing the RG steps described 
in the previous section.

Firstly, to determine the fixed point structure of the flow equations 
and to determine the Kondo temperature 
we continued the RG even  after reaching   the first 
excited state (where the dynamics of the defect must be entirely frozen due to 
the energy splitting) and verified that then the couplings indeed 
flow to the two-channel Kondo fixed point as expected.

The leading logarithmic  scaling equations for many levels  were investigated 
by Aleiner {\it et al.} in the tunneling regime \cite{10}:  
They correspond  to the first two diagrams in Fig.\ref{diagrams}
and  to the second order terms in Eq.~(\ref{renorm_eq}).
Aleiner {\it et al.}  have shown that in leading logarithmic order
the logarithmic contribution of the  excited states 
cancels out due to a sum rule, related to the approximate completeness 
of the defect wave functions.  This reduces the effective cutoff to the 
order of the Debye temperature and thus  the essential renormalization of the couplings 
occurs in the region where only few excited states remain active.

\begin{figure}
\begin{center}
\epsfxsize7.0cm
\epsfbox{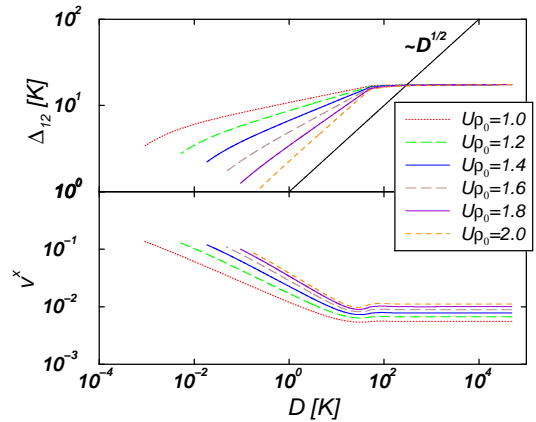}
\end{center}
\caption{\label{coupling_flow} 
The RG flow of $\Delta_{12}$ and $v_x$ in  next-to-leading
logarithmic order for $\alpha=0.4$, $d=0.5{\rm \AA}$ and 
$V_B=100{\rm K}$. Only curves corresponding to $\varrho_0 U \le 2.0$
are shown, since for larger   $\varrho_0 U$ the weak coupling approximation is not valid.
}
\end{figure}

We find that this is also true for the renormalization of the tunneling rate. 
Fig.\ref{coupling_flow} shows the renormalization
of the couplings, $v_x$, and the tunneling rate, $\Delta_{12} \equiv {\tilde \epsilon}_2
-{\tilde \epsilon}_1$,
computed by solving  leading and next-to-leading logarithmic scaling equations 
numerically. Though we could not find a simple sum rule similar to that of Ref.~\cite{10}, 
we found that   both $v_x$ and $\Delta_{12}$  remain unrenormalized  
throughout a large energy region, down to an energy scale
of the order of the Debye temperature $\sim \omega_D$. This is in agreement with 
the 'adiabatic conduction band' picture of Kagan and Prokof'ev,\cite{Kagan} who argued 
that, since the tunneling takes place during a time scale $\sim 1/\omega_D$, electrons 
with energies larger than $\omega_D$ follow the defect instantaneously and 
therefore do not contribute to the logarithmic singularities summed up by the renormalization group.
Therefore   the drastic changes in the effective couplings appear  only when a few 
levels remain.

For typical bare values of the splitting $\Delta\equiv \Delta_{21}\equiv
E_2-E_1 \ge 1 {\rm K}$, the renormalization of  $\Delta$ turned out to be
less than about $25\%$.
This must be  opposed to the case  where only 
two levels are kept from the beginning, and therefore a strong power-law renormalization 
occurs even at high energies \cite{8}.  Our results agree with those obtained by means of an
 adiabatic approximation where the renormalization occurs only when the cutoff 
is already in the range of the few lowest energy levels \cite{Kagan}. 

\begin{figure}
\begin{center}
\epsfxsize7.0cm
\epsfbox{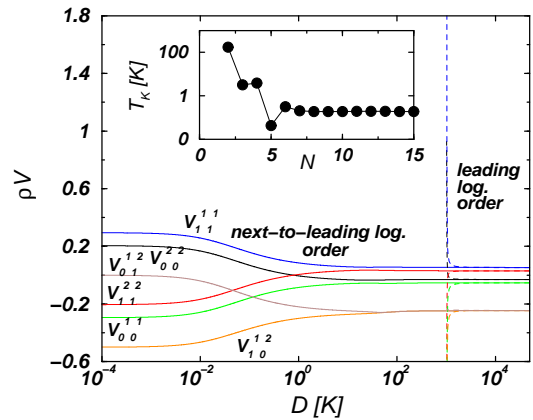}
\end{center}
\caption{\label{TK_numberofstates}
The RG flow of the couplings in leading logarithmic order and
in next-to-leading logarithmic order. Inset:
Kondo temperature as a function of the number of defect states,
for $U\varrho_0=2.1$, $d=0.5{\rm \AA}$ and $V_B=100{\rm K}$ .
}
\end{figure}

It is very instructive, how $T_K$ depends on the number of defect states
(see Fig.\ref{TK_numberofstates}). 
Such a calculation has been first performed by Aleiner {\it et al.}\cite{10}. 
The first few states  result in a strong increase in $T_K$, but then 
$T_K$ suddenly drops, and  the states $n\geq 6$ practically don't produce any 
further changes. In the following only the 'saturation values' of $T_K$
will be reported that we obtained by eventually continuing the RG flows below
the first excited state if $\Delta_{21}>T_K$.

\begin{figure}
\begin{center}
\epsfxsize7.0cm
\epsfbox{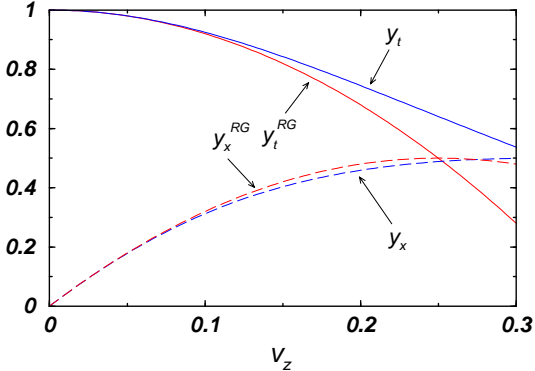}
\end{center}
\caption{\label{fig:exponents}
Comparison of the exact scaling exponents $y_t$ and $y_x$ with their 
approximate values $y^{RG}_t$ and $y^{RG}_x$.} 
\end{figure}

It is very important to determine the range of validity of the perturbative 
scaling analysis we use. Fortunately, there is a few non-perturbative results available 
for the TLS model that can be used to achieve this goal. In the limit 
of small $\Delta_{12}$ and $v_x$ the scaling equations take a simple 
form \cite{VladZimZaw}:
\begin{eqnarray}
{{\rm d} t \over {\rm dx}} & = & y_t\;
t\;, 
\\
{{\rm d} v_x \over {\rm dx}} & = & y_x\;
v_x\;,
\end{eqnarray}
where the scaling exponents $y_t$ and $y_x$ of the dimensionless tunneling 
(splitting) $t\equiv \Delta_{12}/D$ and assisted tunneling $v_x$, 
 only depend  on the coupling $v_z$ in Eq.~(\ref{eq:vz}):
\begin{eqnarray}
y_t &=& 1 - 8 \left({\delta \over \pi}\right)^2 \;,\\
y_x &=& 4 \left({\delta \over \pi}\right) - 
8 \left({\delta \over \pi}\right)^2 \;,\\
{\delta} &= & {\rm artan}(\pi \;v^z)  \;.
\end{eqnarray}
In the multiplicative RG scheme  these non-perturbative exponents are replaced by the 
following approximate exponents: 
\begin{eqnarray}
y^{RG}_t &=& 1 - 8 v_z^2 \;,\\
y^{RG}_x &=& 4 v_z - 
8 v_z^2  \;.
\end{eqnarray}
These approximate exponents are compared to the exact ones in Fig.~\ref{fig:exponents}: 
The exponent $y_x$ of the assisted tunneling is surprisingly well approximated by 
$y^{RG}_x$ in the range $v_z < 0.25$, and therefore we expect that the estimate of the 
Kondo temperature is reliable in the range $v_z \le 0.25-0.3$. However, $y^{RG}_t$ underestimates 
the value  of $y^{RG}_t$ and therefore overestimates the renormalization of $\Delta_{12}$, for which the 
RG results should be trusted only in the range $v_z \le 0.22$.

\begin{figure}
\begin{center}
\epsfxsize6.0cm
\epsfbox{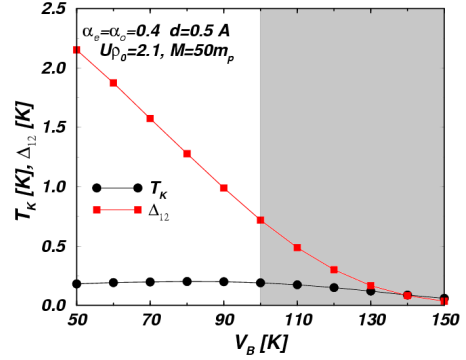}
\end{center}
\caption{The Kondo temperature as a function of 
the height of the central barrier, for $\alpha=0.4$ and $M=50m_p$. 
The renormalized level splittings are plotted as well.
The shaded part of the figure indicates the region where the estimate of 
the renormalized $\Delta_{12}$ is unreliable.}
\label{fig:Vfug}
\end{figure}

Fig.~\ref{fig:Vfug} shows the barrier height dependence of the Kondo temperature: 
 $T_K$  increases  with increasing barrier height up to a certain value of
$V_B$ and then it drops suddenly, while  the splitting $\Delta_{21}$ 
decreases continuously. Initially, increasing the barrier height concentrates the 
wave functions of the first 
two states more and more around the potential minima and this results in an 
increase of the coupling $v_z$, and thus a gradual increase in $T_K$. 
This tendency is, however, suddenly reversed once the barrier reaches  the 
first level, and one enters the tunneling regime
($V_B \ge 150 {\rm K}$ for the parameters of Fig.~\ref{fig:Vfug}): 
There the assisted tunneling $v_x$, responsible for the 
generation of the Kondo effect, decreases exponentially, and both $\Delta_{12}$ and $T_K$ 
decrease dramatically.

\begin{figure}
\begin{center}
\epsfxsize6.0cm
\epsfbox{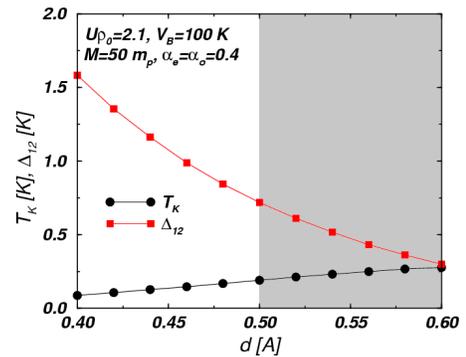}
\end{center}
\caption{ 
The Kondo temperature and the renormalized level splittings as a function of 
the total width $d$ of the potential well
for $\alpha=0.4$ and $M=50m_p$.
The shaded range of the figure indicates the region where the estimate of 
the renormalized $\Delta_{12}$ is unreliable.}
\label{fig:dfug}
\end{figure}

$T_K$ is also  very sensitive to the width of the entire potential well $d$. In Fig.~\ref{fig:dfug}
we show the $d$-dependence of $T_K$ for a defect that has a relatively low barrier height
and is not in the tunneling regime. As $d$ is decreased from $d=0.5\AA$, 
the energy levels are shifted to higher values and
also the initial values of the  coupling $v_z$  decrease while $v_x$ does not change 
dramatically. Thus the width of the well is a fundamental parameter and the Kondo effect can only 
occur if the room for the
particle considered is anomalously large in one direction. For defects in the tunneling range
$T_K$ decreases with increasing $d$ since the bare value of $v_z$ only slightly increases while 
the assisted tunneling $v_x$ is exponentially suppressed in the tunneling regime. 

\begin{figure}
\begin{center}
\epsfxsize6.0cm
\epsfbox{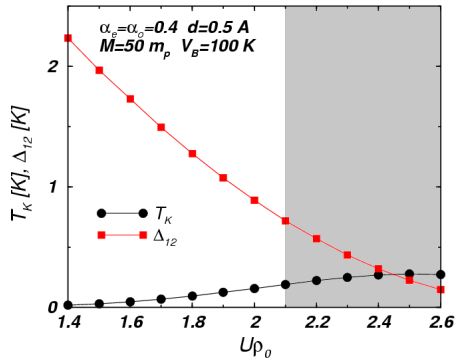}
\end{center}
\caption{
The Kondo temperature and renormalized level
splittings as a function of the scattering strength $U$. 
The shaded part of the figure indicates the region where the RG estimate of 
the renormalized $\Delta_{12}$ is unreliable.}
\label{fig:Udep}
\end{figure}

In Fig.~\ref{fig:Udep} we show the dependence on the strength of the defect-electron 
interaction $U$. Again, $T_K$ increases continuously with increasing $U$, 
since  all couplings generating the Kondo effect increase. 
On the other hand, The renormalized $\Delta$ 
 decreases continuously since the scaling exponent $y_t^{RG}$ also decreases with 
increasing $U$. 

Finally, let us discuss the electron-hole asymmetry dependence of 
$T_K$. For $U_0=0$, $\varrho_{l=0} = \varrho_{l=1}$
 and $\alpha= \alpha_0 = \alpha_1$ we find no significant 
$\alpha$-dependence of the renormalized splitting and $T_K$. 
In general, however,  the local density of states is different for 
the $l=0$ and $l=1$ channels and thus both $\varrho_0$ and $\alpha$ depend
on $l$.  A non-zero  potential scattering term $U_0 \ne 0$ in 
Eq.~(\ref{H1}), {\em e.g.},  would have  a twofold effect: (1) It 
would change the local  density of states in the $l=0$ scattering channel, 
and  (2) would generate a different electron-hole symmetry breaking for 
the $l=0$ and $l=1$ channels \cite{14}.  
 [Note that 
in Ref.~\onlinecite{comment} $U_0$ has not been set to zero and has been 
treated inappropriately: Instead of incorporating it in the 
parameters $\varrho_0$ and $\alpha_0$ we included it in the RG 
equations as a coupling constant.  Though the results obtained are 
qualitatively similar to the ones we obtained below, this procedure is 
wrong, since mass terms like $U_0$ 
have to be treated always  differently from 
the coupling constants in the multiplicative RG \cite{14}.]

\begin{figure}
\begin{center}
\epsfxsize6.0cm
\epsfbox{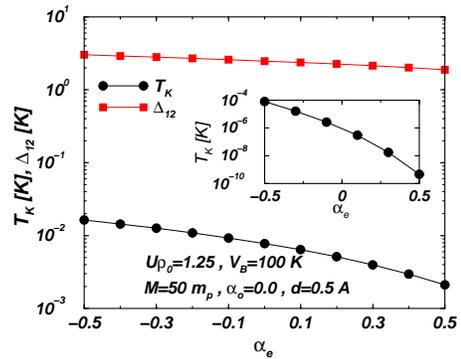}
\end{center}
\caption{
{
Electron-hole asymmetry dependence of $T_K$ and the splitting
$\Delta_{12}$. Inset: the effect is more dramatic for very small $T_K$'s.} 
} 
\label{fig:alpha}
\end{figure}

It is easy to generalize our previous calculations to the case
$\alpha_0\ne \alpha_1$ and $\varrho_{l=0}\ne 
\varrho_{l=1}$: The $l$-dependent densities of state
can be treated by defining the fermion fields slightly differently.
However, for $\alpha_0\ne \alpha_1$ the electron-hole symmetry breaking 
generates strong non-logarithmic corrections  to the bare coupling 
constants. In Fig.~\ref{fig:alpha} we plotted 
the dependence of $T_K$  on $\alpha_0$ while keeping $\alpha_1$ zero
and $\varrho_{l=0}=\varrho_{l=1}$.
We find that  for typical parameters we used earlier $T_K$ can change
about an order of magnitude. This renormalization 
can have even more dramatic effects for very small $T_K$'s where 
$T_K$ can change several orders of magnitude due to changes in the value of the 
electron-hole asymmetry (see inset of Fig.~\ref{fig:alpha}).

Let us conclude this Section with a general observation:\\
Increasing the mass (the $M/m_p$ ratio), the energies of the levels are
scaled down. If the central barrier is similarly reduced then $T_K$ becomes smaller 
by the same ratio and thus $\Delta/T_K$ is not affected, as the dependence
on the high energy cutoff $D$ is weak. \\

%

\section{Conclusions}

In this paper we performed a thorough scaling analysis of a dynamical defect coupled to 
the conduction electrons, by taking into account all the excited states of the 
defect, and  constructing the next to leading logarithmic scaling 
equations. We also discussed how to take into account large non-logarithmic terms that 
renormalize the bare parameters of the theory. In the perturbative regime
our calculations confirmed   the results of Ref.~\onlinecite{10}: The electron-defect 
couplings  remain unrenormalized down to an energy scale of the order of 
the Debye temperature $D\sim \omega_{\rm Debye}$ due to a cancellation pointed 
out in Ref.~\onlinecite{10}. 

In the leading logarithmic approach of Ref.~\onlinecite{10} the energy levels of 
the impurity are unrenormalized, and the above next to leading logarithmic 
analysis is needed to compute the renormalization of the defect states. 
We found that, similar to the defect-electron coupling, the splitting of the first two defect 
states remains unrenormalized down to $\omega_{\rm Debye}$. In contrast to  the expectations 
of Ref.~\cite{Moustakas}, where the generation of a large splitting has been predicted, we 
found that the renormalization of the defect states remains small if one takes into 
account all defect states. 

Our results perfectly agree with the adiabatic picture of Kagan and Prokof'ev \cite{Kagan}:
Electrons with excitation energies $|\epsilon| > \omega_{\rm Debye}$ follow the defect instantaneously, 
and therefore the role of the excited states is only reducing the effective electronic cut-off from the 
order of the Fermi energy down to the energy scale $ \omega_{\rm Debye}$.

We also analyzed the range of validity of our approach by using some analytical  results
in the strong coupling regime. We found that the perturbative RG predicts the 
Kondo temperature correctly for surprisingly large couplings, however, 
it tends to overestimate the renormalization of the splitting $\Delta$. 
 
We solved the RG equations for initial couplings  obtained 
by changing various parameters of the double well potential. 
We always found a region where $T_K$ and the renormalized  $\Delta_{12}$ 
were comparable and therefore one should be able to observe the Kondo anomaly
in various measurements. Our calculations show that to have 
$T_K\sim\Delta_{12}$ in the measurable range 
one needs a defect that ({\em i}) has anomalously large room to move in one direction,
({\em ii}) is close to but not yet in the tunneling regime ({\em iii}), which has a large 
scattering strength, $U\varrho_0\geq 2.5$, implying  resonant 
scattering on the particle. The best candidates are therefore atoms with resonant 
d or f-scattering at the Fermi energy or possibly small groups 
of atoms\cite{20}. Thus dynamical local defects with special
realistic model parameters can explain the Kondo-like anomalies observed in 
some experiments for $T>T_K$\cite{1,3,6}. Note that in amorphous materials the 
positions of the defect $f$- or $d$-orbitals have usually a distribution, and therefore 
many of the defects may have resonant scattering at the Fermi energy. 

However, to explain the two-channel Kondo scaling reported in Ref.~\cite{Buhrman},  
one would need a renormalized $\Delta_{12}$ that is less than $T_K$ at the energy 
scale $D\sim T_K$.  Unfortunately, 
our calculations are only of logarithmic accuracy, and therefore we cannot 
decide if the ratio $\Delta_{12}/T_K$ is small enough to display a clear two-channel Kondo 
scaling. In the perturbative regime (i.e. the regime where where our perturbative RG works), our 
results seem to indicate that 
$T_K$ can be comparable to, but is somewhat  smaller than $\Delta_{12}$, 
and is thus too small to result in an observable  two-channel Kondo  
behavior.  However, one can  prove both analytically and with NRG calculations the existence 
of a  regime  {\em outside} the range  of perturbative RG \cite{resonant} where $T_K$ can 
be larger than $\Delta_{12}$.   This is also indicated by our perturbative results.  

We have to emphasize that for a defect with resonant scattering in a disordered environment 
many of our approximations (free electron model, simple delta scattering on the defect, usage of a 
local effective potential etc.)
are questionable, and the estimates of the various couplings are therefore inadequate.
Also, in reality, the dynamical defect is not formed by a single atom, but 
it is probably rather a  group of atoms (dislocation, e.g.) that is responsible for 
the defect dynamics. 
Though our general conclusions probably do not depend on these approximations, 
$T_K$, {\em e.g.}, depends exponentially on the coupling constants and 
more sophisticated models/calculations would be needed to give a quantitative estimate 
of $T_K$ and the ratio $\Delta_{12}/T_K$. 

It remains an open question whether for larger couplings ({\em i.e.}, defects with resonant 
scattering at the Fermi energy) a larger $T_K\sim 5-10 {\rm K}$ can be reached, 
as required to explain the experiments for Refs.~\onlinecite{1,2,3}. 

We are grateful to B. Altshuler, I. Aleiner, Yu. Kagan, and N. Prokof'ev for valuable
discussions.  We benefitted from the  discussions and hospitality
of J. von Delft at Munich and the support of the Humboldt Foundation
(A. Z.). This research has been supported by Hungarian Grants 
No. T03038243, T034243, T038162,  F030041, 
TS040878, 
NSF Grant No.  
DMR 99-81283,
and EU RTN No. RTN2-2001-00440. 


\begin{references}
\bibitem{footnote} The fixed point value of the couplings in the next to 
leading logarithmic order is $v^x_{\rm fp} = v^y_{\rm fp} = 
v^z_{\rm fp} = 0.25$.
\end{references}
\end{document}